\documentclass[preprint,prc,aps,showpacs,showkeys,superscriptaddress]{revtex4} 
\usepackage{ae,aecompl}
\usepackage[T1]{fontenc}
\usepackage[latin9]{inputenc}
\usepackage{amsmath}
\usepackage{float}
\usepackage{amssymb}
\usepackage{graphicx}
\usepackage{multirow}
\usepackage{mathtools}
\usepackage{pifont}

\makeatletter

\pdfpageheight\paperheight
\pdfpagewidth\paperwidth

\newcommand{\noyau}[3]{\prescript{#2}{#3}{\mathrm{#1}}}
\newcommand{\ee}[1]{$\times 10^{#1}$}

\begin{document}

\title{Zero degree measurements of $\noyau{C}{12}{}$ fragmentation at 95~MeV/A on thin targets}

\author{J. Dudouet}
\author{M. Labalme}
\author{D. Cussol}
\affiliation{LPC Caen, ENSICAEN, Universit\'e de Caen, CNRS/IN2P3, Caen, France}
\author{C. Finck}
\affiliation{Institut Pluridisciplinaire Hubert Curien Strasbourg, France}
\author{R. Rescigno}
\author{M. Rousseau} 
\affiliation{Institut Pluridisciplinaire Hubert Curien Strasbourg, France}
\author{S. Salvador}
\affiliation{LPC Caen, ENSICAEN, Universit\'e de Caen, CNRS/IN2P3, Caen, France}
\author{M. Vanstalle}
\affiliation{Institut Pluridisciplinaire Hubert Curien Strasbourg, France}

\date{\today}

\begin{abstract}
During therapeutic treatment with heavy ions like carbon, the beam undergoes nuclear fragmentation processes and secondary light charged particles, in particular protons and alpha particles, are produced. To estimate the dose deposited into the tumors and the surrounding healthy tissues, an accurate prediction on the fluencies of these secondary fragments is necessary. Nowadays, a very limited set of double differential carbon fragmentation cross sections are being measured. An experiment has been realized in 2011 at GANIL to obtain the double differential fragmentation cross sections for a 95~MeV/A carbon beam on different thin targets at angles from 4 to 43$^\circ$. In order to complete these data, a new experiment has been performed on September 2013 at GANIL to measure the fragmentation cross section at zero degree for a 95~MeV/A carbon beam on thin targets. In this work, the experimental setup will be described, the analysis method detailed and the results presented.
\end{abstract}



\maketitle

\section{Introduction}

The use of carbon ion beams in hadrontherapy to treat cancerous tumors is motivated by the highly localized dose distribution. The high carbon mass (compared to proton) leads to a smaller angular scattering and a higher dose deposition at the end of the radiation range ({\it i.e.} at the Bragg peak). Moreover, the biological efficiency, which is strongly correlated to the linear energy transfer (LET), is higher for carbon ions in the Bragg peak region compared to protons. Carbon ions allow thus to better targeting the tumor while preserving the surrounding healthy tissues. However, the physical dose deposition is affected by nuclear reactions of the ions along their penetration path in human tissues~\cite{Schardt96}. As a consequence, the number of incident ions reaching the tumor is reduced, for instance, up to 70\%  for 400~MeV/A $^{12}$C in water. The carbon beam fragmentation in the human body leads to a mixed radiation field composed of lighter fragments of higher ranges and of broader angular distributions with respect to the primary ions. These lighter fragments have different relative biological effectivenesses (RBE) which contribute to the deposited dose all along the carbon path. These effects result in a new spatial dose distribution, particularly on healthy tissues. This must be taken into account for the evaluation of the biological dose~\cite{Scholz00} by accurately evaluating the fragmentation processes.

Simulation codes are used to compute the transportation of ions in matter but the constraints on nuclear models and fragmentation cross sections at therapeutic energies (up to 400 MeV/A) are not yet sufficient to reproduce the fragmentation processes with the required accuracy for clinical treatments~\cite{Bohlen10,Braunn12,Catane12}. Nuclear cross sections are critical inputs for these simulation frameworks. In particular, there is a lack of experimental cross section data available for light ions on light targets in the energy range from 30 to 400 MeV/A. These experimental data are necessary to benchmark Monte Carlo codes for their use in hadrontherapy.

To improve the models and reach the accuracy required for a reference simulation code for hadrontherapy ($\pm$3\% on dose value and $\pm$2 mm spatial resolution), several experiments on thin targets have been planned. An experiment was performed at 62 MeV/A on carbon target in Catania~\cite{Catane12}. Another one was performed at 400 MeV/A on carbon target at the Gesellschaft f\"ur SchwerIonenforschung (GSI, Germany) by the FIRST collaboration~\cite{FIRST12}. A third experiment was performed by our collaboration on May 2011 at the Grand Acc\'el\'erateur National d'Ions Lourds (GANIL, France) to study carbon reactions on C, H, O, Al and $^{nat}$Ti targets at 95 MeV/A~\cite{Dudouet13b}. These experiments have shown that the fragment yields are largely dominated by projectile fragmentation and that the angular distributions are highly focused toward the small angles. An in beam measurement (thereafter denoted as zero degree measurement) should thus be the most constraining information on nuclear models. However,
 no zero degree carbon fragmentation cross section are available in the literature and databases. In view of this remark, a second experiment at 95~MeV/A has been performed by our collaboration on September 2013 at GANIL and will be described in this work. In the following part, the experimental setup will be detailed. The analysis method used will then be described and finally, the experimental results will be presented.

\section{Experimental setup}

\begin{figure}[!ht]
\includegraphics[width=1\linewidth]{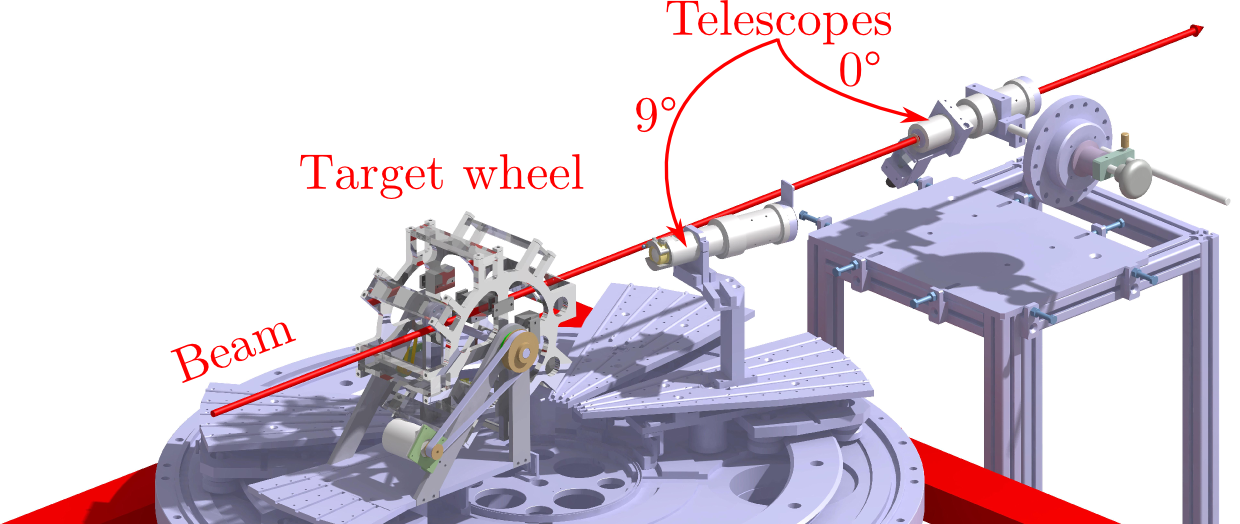}
\caption{(Color online) Schematic view of the experimental setup.}
\label{fig:Setup}
\end{figure}

The experimental setup is similar to the one used in the previous experiments performed by our collaboration~\cite{Braunn11,Dudouet13b}. The ECLAN reaction chamber has been used. The experimental setup is represented in the schematic view of Fig.~\ref{fig:Setup}. A carbon beam at 94.98(9)~MeV/A was used to produce fragmentation processes on different thin targets [C, CH$_2$, Al, Ti]. The targets were placed at the center of the reaction chamber, on a rotating target handler. The target area densities were of about 200~mg.cm$^{-2}$. In order to increase the fragmentation process statistics, the targets were made four times thicker than in the previous experiment. The targets characteristics are detailed in Table~\ref{tab:targets_char}.

Concerning the charged particles detection, two  telescopes have been used. A first one was placed in the reaction chamber at 9$^\circ$ with respect to the beam direction. This telescope has been used to cross-check the results of this experiment with the 9$^\circ$ values of the previous one. The second telescope was placed in the beam axis, at a distance of 768(5)~mm behind the targets. The two telescopes were made of a stack of two silicon detectors followed by a CsI(Tl) crystal scintillator. The detectors properties are presented in Tab.~\ref{tab:telescope} for memory.

In order to reduce pile-up events due to the low response time of the CsI crystal, low intensities (10$^3$-10$^4$ pps) were needed. But even for such intensities, the probability to detect two or more incident ions in successive beam bunches in a time shorter than the CsI output signal duration (10~$\mu$s) is quite important (5-10\%). Thanks to the homemade digital acquisition, FASTER~\cite{FASTER}, the signal shapes obtained were analyzed to detect events in which the signals of different particles are piled-up. An algorithm analyses the signals and if two or more maxima are found on one signal, the event is labeled as ``pile-up event''. The pile-up events have been removed from the experimental data.

To determine the number of incident ions, a beam monitor has been placed before the reaction chamber. This beam monitor is exactly the same than the one used in the previous experiment. It was based on the measurement of fluorescence X-rays emitted by a thin Ag foil (7~$\mu$m thick) which is set perpendicular to the beam axis. X-rays are detected by means of a Si(Li) detector located at $90^\circ$ with respect to the beam direction. More details on this beam monitor are available in Dudouet {\it et al.}~\cite{Dudouet13a}.

During this $0^\circ$ experiment the number of incident ions could also be estimated by the zero degree telescope itself, since it was located in beam. This redundancy between these two measurements of the number of incident ions  with a precision better than 1\% allowed to validate the beam monitor calibration at low intensities ($\sim 10^3$ pps). A precise study of the beam monitor calibration was then performed at higher intensities. It revealed that the linearity of the plastic scintillator used to calibrate the beam monitor was lost for events containing more than five incident ions per beam bunch. Only calibration points at lower intensities were therefore used. The main consequence of this observation concerns the experiment performed in 2011. In this experiment~\cite{Dudouet13a,Dudouet13b}, a non linearity of the beam monitor has been suspected for the highest intensities. However, without certainty, the beam monitor calibration was made keeping a calibration point for which the number of incident ions per bunch was higher than 5. Following the previous observation, this point has been removed from the calibration of the beam monitor of the 2011 experiment. The consequence is an increase of about 5\% of the estimation of the number of incident ions of this experiment. Although this error was contained in the experimental error bars, the choice has been made to correct the data presented in Dudouet {\it et al.}~\cite{Dudouet13b}. These corrected data have been updated on the website \url{http://hadrontherapy-data.in2p3.fr} and will be used in this article.

\begin{table}[!ht]
\begin{center}
{\setlength{\tabcolsep}{3mm}
\begin{tabular}{c c c}
\hline
\hline
Target		&th (in mm)	 & $\rho \times$th (in mg~cm$^{-2}$) 	\\
\hline
C 		&  1.0		 & 176(2)				\\

CH$_2$ 		&  2.0		 & 179(2)				\\

Ti 		&  0.5		 & 226(2)				\\

Al 		&  0.7		 & 177(2)				\\
\hline
\hline
\end{tabular}}
\end{center}
\caption{Target characteristics. The targets thicknesses are labeled $th$ and their densities are labeled $\rho$.}
\label{tab:targets_char}
\end{table}

\begin{table}[!ht]
\begin{center}
{\setlength{\tabcolsep}{3mm}
\begin{tabular}{c c c c c}
\hline
\hline
		& Thin Si	& Thick Si	& CsI 		& $\Omega$ (msr)	\\
\hline
Thicknesses 	& 146~$\mu$m 	& 1~mm		& 12~cm        	& 0.51(3)		\\
\hline
Diameter	& 1.954~cm	& 1.954~cm	& 3~cm		&			\\
\hline
\hline
\end{tabular}}
\end{center}
\caption{Geometrical properties of the detectors.}
\label{tab:telescope}
\end{table}

\section{Analysis}

Two different analyses were needed for this experiment. The first one, used to study the 9$^\circ$ telescope, is exactly the same that the one used in our first experiment described in Dudouet {\it et al.}\cite{Dudouet13a}. However, for the zero degree telescope, the large amount of projectiles interactions with the detectors makes this identification method impossible to apply. The analysis method that was used will now be described.

\begin{figure*}[!ht]
\includegraphics[width=1\linewidth]{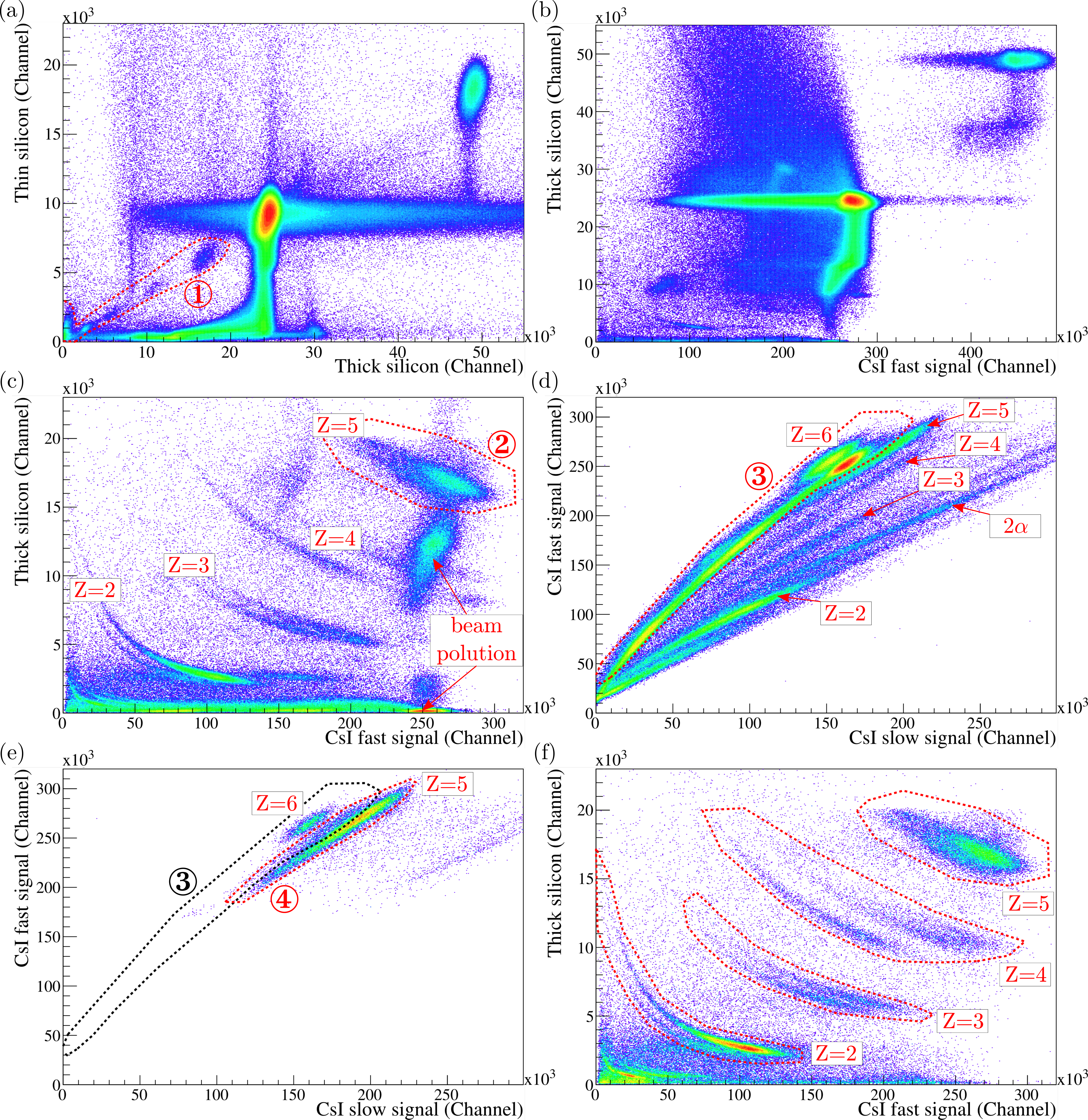}
\caption{(Color online) (a,b) $\Delta$E/E maps without any selection. (c) $\Delta$E/E map cleaned-up from \ding{172}. (d) CsI Fast/Slow map cleaned-up from \ding{172}. (e) CsI Fast/Slow map cleaned-up from \ding{172} and \ding{174}. (f) Final $\Delta$E/E map resulting of the combination of graphical cut \ding{172}-\ding{175}.}
\label{fig:Analysis}
\end{figure*}

Two identification methods were combined in this work. The first one is a $\Delta$E-E identification. It consists in representing the energy lost in a first detector as a function of the one lost in a second detector. Two $\Delta$E-E maps are then obtained: a first one representing the energy lost in the thin silicon detector as a function of the one lost in the thick silicon detector (cf.~Fig.~\ref{fig:Analysis}(a)) and a second one representing the energy lost in the thick silicon detector as a function of the one lost in the CsI crystal (cf.~Fig.~\ref{fig:Analysis}(b)). 

The second identification method used relies on a CsI scintillation pulse shape analysis. It has been shown that the CsI(Tl) light output could be described to a good approximation by a sum of two exponential functions associated with a fast and a slow component~\cite{Ben89}. Using the digital acquisition, which treats the CsI signals with a charge to digital converter (QDC), two time windows were chosen for integrating the scintillation light. The first window was called ``fast'' [0:500~ns] and the second one, ``slow'' [1~$\mu$s:5~$\mu$s]. By representing the fast component of the signal versus the slow one, the so called CsI(fast/slow) identification map was obtained, on which each isotope can be identified (cf.~Fig.~\ref{fig:Analysis}(d)).

Fig.~\ref{fig:Analysis}(a,b) represents the $\Delta$E-E maps without any correction. The statistic is dominated by the incident carbon ions (coordinates (24000;9000) on Fig.~(a) and (280;24000) on Fig.~(b)). A second peak, located two times higher in energy, corresponds to the detection of two incident carbon ions in the same bunch. The interactions between the incident ions and the whole detectors (including the sensitive and mechanical parts) generate lots of pollutions on the $\Delta$E-E maps (horizontal and vertical trails). The fragments, stopped in the CsI, need to be identified on the $\Delta$E$_\text{Thick Si}$-E$_\text{CsI}$ map (b), but are covered by the pollution due to beam interactions in the detectors. 

In order to clean this $\Delta$E-E map, a combination of several graphical selections which will now be described has been needed. At zero degree, the emitted fragments are mainly produced at velocities close to the beam velocity. The fragments are almost always passing through the thick silicon (e.g. punch through particles): selected by the graphical selection \ding{172} on the $\Delta$E$_\text{Thin Si}$-E$_\text{Thick Si}$ map (a). It has to be noted that, as the particles are selected from the punch through lines, the experimental energy thresholds in this experiment are determined as the required energy for a given isotope to pass through the two silicon detectors (1150~$\mu$m). These energy thresholds are mentioned in Table~\ref{tab:EThr} for the isotopes that have been identified (as explained in the following).

\begin{table}[!ht]
\begin{center}
{\setlength{\tabcolsep}{2mm}
\begin{tabular}{l c c c c}
\hline
\hline
Isotope			& $^4$He	& $^6$Li	& $^7$Li	& $^7$Be	 	\\	
\hline
E$_{\text{th}}$ (MeV/A)	& 13.1		& 16.6		& 15.2		& 21.0			\\
\hline
\hline
\end{tabular}}
\caption{Experimental energy thresholds for the different isotopes that were identified in this work.}
\label{tab:EThr}
\end{center}
\end{table}

By applying the graphical selection \ding{172} used to select the punch-through particles, events coming from the beam are consequently removed. Fig.~\ref{fig:Analysis}(c) represents the events in selection \ding{172} on the $\Delta$E$_\text{Thick Si}$-E$_\text{CsI}$  map. Although most of the pollution was removed, the $\Delta$E$_\text{Thick Si}$-E$_\text{CsI}$ map is still not cleaned from all pollution sources. The remaining pollution comes from incident carbon ions that have interacted with the supports of the silicon detectors (metal frame used to hold the sensitive area of the silicon detector). For such interactions, the induced charges are not well collected, resulting in output signals of random shapes and amplitudes. Thanks to a larger diameter of the CsI crystal, compared to the one of the silicon detectors, even if they are not passing through the sensitive area of the silicon detectors, these carbon ions can still be identified in the CsI(fast/slow) map (cf.~Fig.~\ref{fig:Analysis}(d)). The selection \ding{173}, made on the CsI(fast/slow) map, is used to clean the pollution created by these incident carbon ions on the $\Delta$E$_\text{Thick Si}$-E$_\text{CsI}$ map. However, carbon and boron isotopes are not well separated on the CsI(fast/slow) map. Such a selection will thus contain events corresponding to boron isotopes that need to be kept. To solve this problem, a graphical selection \ding{174} was made for boron isotopes on the $\Delta$E$_\text{Thick Si}$-E$_\text{CsI}$ map. By looking at this selection on the CsI(fast/slow) map, the boron isotopes can now be separated from the carbon ions (selection \ding{175} on Fig.~\ref{fig:Analysis}(e)). To obtain a cleaned-up $\Delta$E$_\text{Thick Si}$-E$_\text{CsI}$ map, any event in the selection \ding{173} will be remove unless this event is a real boron isotope (event contained in selections \ding{174} and \ding{175}). Finally, once all selections were applied, a cleaned-up $\Delta$E$_\text{Thick Si}$-E$_\text{CsI}$ map was obtained on which the different isotopes were well separated (cf.~Fig.~\ref{fig:Analysis}(f)).

This analysis allowed to identify particles from Z=2 to Z=5 (graphical selections on Fig.~\ref{fig:Analysis}(f)). It was necessary to impose that the thin silicon detector was hit to avoid other pollutions due to the beam interactions that could not be cleaned. The drawback of this condition is that a part of the high energy Z=1 particles were below the experimental energy thresholds of the thin silicon detectors. For this reason, the cross sections of Z=1 were not measured in this experiment. Due to the small statistics obtained in this experiment, only the most produced isotopes were separated by graphical selections.

Regarding the detectors energy calibration, a similar method than the one described in Dudouet {\it et al.}~\cite{Dudouet13a} was used. Some reference events for which the energy is known are needed. The detection of the beam gives us four calibration point (one per target). Indeed, as the beam energy and the targets and silicon detectors thicknesses are well known (<1\%), the energy deposited by the incident carbon ions in the two silicon detectors can be calculated. Moreover, in order to obtain calibration points to constrain low energies, a three-alpha source ($\noyau{Am}{241}{}$, $\noyau{Cu}{244}{}$, $\noyau{Pu}{238}{}$) was used. Once the two silicon detectors were calibrated, the CsI energy was determined from the silicon detectors calibration by energy-loss calculation as the remaining energy of the particle after passing through the two silicon detectors.

\section{Experimental results}

The results of the cross-check measurement at 9$^\circ$ with our previous experiment will firstly be presented and the zero degree results will then be detailed. Some comparisons with data at larger angles obtained in the previous experiment will finally be done.

\subsection{9$^\circ$ cross sections cross-check}

As explained in the previous part, a telescope was placed at an angle of 9$^\circ$ with respect to the beam direction. This measurement was done in our previous experiment~\cite{Dudouet13b}. Due to the low intensities used in this current experiment, the statistics obtained for the 9$^\circ$ telescope are very small. As a consequence, it was not possible to obtain the cross sections for each isotope. The comparisons will thus only be made on the different Z values. These comparisons are reported in Table~\ref{tab:cross-check} for Z=1 to Z=3 and for the carbon, aluminum and titanium targets. The lines labeled ``2013'' in Table~\ref{tab:cross-check} are the results of this new experiment and the ones labeled ``2011'' refer to the previous experiment. The indicated relative errors are calculated as the deviation from the average of the two values.

\begin{table*}[!ht]
\begin{center}
\begin{tabular}{l c c c c}
\hline
\hline
\multicolumn{1}{l}{\multirow{2}{2cm}{Target}} & \multicolumn{3}{c}{$\partial\sigma/\partial\Omega$ (b sr$^{-1}$)}\\
\cline{2-4}
	& 	\multirow{1}{2cm}{\centering Z=1}	& 	\multirow{1}{2cm}{\centering Z=2}	& 	\multirow{1}{2cm}{\centering Z=3}	\\
\hline
C 2013 		&4.63(0.69)	&4.53(0.68)	&0.34(0.05)	\\

C 2011		&4.65(0.30)	&4.68(0.47)	&0.35(0.03)	\\

\hline

Rel err (in \%)	& 0.4		& 3.3 		& 2.9 		\\

\hline
\hline

Al 2013 		&7.0(1.1)	&6.23(0.93)	&0.48(0.07)	\\

Al 2011		&6.96(0.45)	&6.31(0.62)	&0.49(0.05)	\\

\hline

Rel err (in \%)	& 0.6		& 1.3 		& 2.0		\\
\hline
\hline

Ti 2013 		&8.9(1.3)	&7.4(1.1)	&0.60(0.09)	\\

Ti 2011		&9.09(0.59)	&7.64(0.75)	&0.59(0.06)	\\

\hline

Rel err (in \%)	& 1.3		& 2.8 		& -1.7 		\\
\hline
\hline
\end{tabular}
\end{center}
\caption{Cross-check between this experiment and the previous one for the 9$^\circ$ telescope.}
\label{tab:cross-check}
\end{table*}

For the different Z values and the different targets, the results of both experiments are in very good agreement (up to 3\% of deviation for the different Z values and for the three targets). This allows us to be very confident in the target properties which are used to determine the number of target nuclei and in the estimation of the number of incident ions which are both used in the cross section calculation and on the evaluation of the systematic errors.

\subsection{Zero degree cross sections}

It has to be noted that due to the active surface of the detectors, a real zero degree measurement is not possible. The center of detector is located in beam, but events are integrated for theta values ranging from zero to a maximum theta value of $\theta_{max}=0.73^\circ$, calculated from the solid angle of the detector. The mean angle of the detector, for a $\partial\sigma/\partial\Omega$ distribution, is 0.36$^\circ$ in the case of a uniform distribution. However, as shown in previous works~\cite{Dudouet13b,Catane12}, the distributions are not uniform but peaked toward 0$^\circ$. The mean value of the zero degree telescope will thus be between 0$^\circ$ and 0.36$^\circ$, depending for each isotope on the slope of the distribution of the angular range covered by the telescope. As the real distribution is not known, and in order to avoid a model dependent correction on this angle value, the choice was made to represent these cross sections at 0$^\circ$, the error made on the angle being included in the 
error bars.

Concerning the uncertainties, the systematic errors were estimated using GEANT4 simulations~\cite{Agostinelli03}. As mentioned in Dudouet {\it et al.}~\cite{Dudouet14}, the INCL++~\cite{Boudard13} model provided in the GEANT4 toolkit best reproduces the angular distributions for the smaller angles (for a 95~MeV carbon beam). For this reason, simulations including the whole experimental setup were performed using this model. The simulated data were analyzed using the same method than the one used for the experimental data and the pile-up events of two alpha particles from the same event in a telescope was treated as described in Dudouet {\it et al.}~\cite{Dudouet13b}. The systematic errors have been estimated at 5-10\% resulting in a total uncertainty on the zero degree cross sections of 10-15\%.

The differential cross sections obtained for the zero degree telescope are represented in Tables~\ref{tab:XsecperZ} and \ref{tab:XsecperFrag}. Tables~\ref{tab:XsecperZ} represents the differential cross sections of fragments production per Z and Table~\ref{tab:XsecperFrag} represents the differential cross sections per isotope for the different targets. As expected, the cross-sections increases with the target mass.

By superimposing these zero degree measurements on the angular distributions obtained in the previous work, we can see that the fragments production is dominated by the small angles (cf.~Figs.~\ref{fig:GEE_AllTarg} and \ref{fig:GEE_AllFrag}). The knowledge of the zero degree cross section is mandatory for an accurate estimation of the fragments production.

\begin{table*}[!ht]
\begin{center}
\begin{tabular}{l c c c c}
\hline
\hline
\multicolumn{1}{l}{\multirow{2}{1.5cm}{Target}} & \multicolumn{4}{c}{$\partial\sigma/\partial\Omega$ (b sr$^{-1}$)}\\
\cline{2-5}
	& 	\multirow{1}{2cm}{\centering Z=2}	& 	\multirow{1}{2cm}{\centering Z=3}	& 	\multirow{1}{2cm}{\centering Z=4}	& 	\multirow{1}{2cm}{\centering Z=5}	\\
\hline
H 	&9.4(2.7)	&1.28(0.38)	&1.31(0.42)	&5.7(1.7)	\\

C 	&30.8(2.9)	&3.99(0.38)	&4.65(0.44)	&18.7(1.7)	\\

Al 	&46.9(4.7)	&5.25(0.51)	&7.42(0.72)	&26.8(2.5)	\\

Ti 	&64.8(6.0)	&7.23(0.70)	&11.3(1.1)	&33.1(3.1)	\\
\hline
\hline
\end{tabular}
\end{center}
\caption{Differential cross sections at 0$^\circ$ per Z and for different targets.}
\label{tab:XsecperZ}
\end{table*}

\begin{table*}[!ht]
\begin{center}
\begin{tabular}{l c c c c}
\hline
\hline
\multicolumn{1}{l}{\multirow{2}{1.5cm}{Target}} & \multicolumn{4}{c}{$\partial\sigma/\partial\Omega$ (b sr$^{-1}$)}\\
\cline{2-5}
& 	\multirow{1}{2cm}{\centering $\noyau{He}{4}{}$}	& 	\multirow{1}{2cm}{\centering $\noyau{Li}{6}{}$}
& 	\multirow{1}{2cm}{\centering $\noyau{Li}{7}{}$}& 	\multirow{1}{2cm}{\centering $\noyau{Be}{7}{}$}\\
\hline
H &8.9(4.2)	&0.75(0.32)	&0.54(0.29)	&0.82(0.32)		\\

C &28.5(4.4)	&2.03(0.32)	&1.96(0.31)	&1.87(0.29)		\\

Al&42.6(6.5)	&2.48(0.40)	&2.77(0.44)	&2.47(0.39)		\\

Ti&57.6(8.8)	&3.47(0.55)	&3.76(0.59)	&3.54(0.56)		\\
\hline
\hline
\end{tabular}
\end{center}
\caption{Differential cross sections at 0$^\circ$ per isotope and for different targets.}
\label{tab:XsecperFrag}
\end{table*}

\begin{figure}[!ht]
\includegraphics[width=\linewidth]{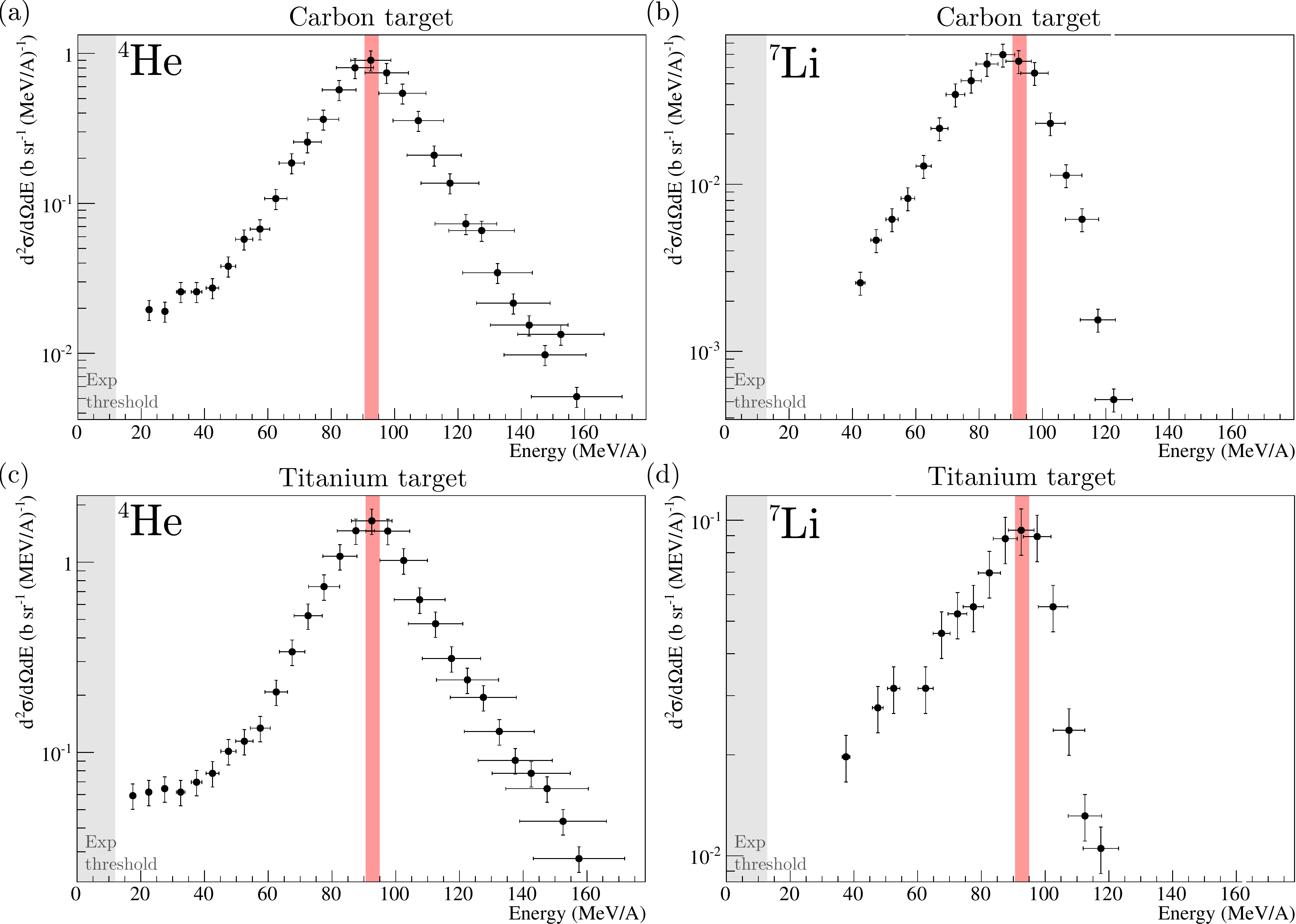}
\caption{(Color online) Energy distribution of $\noyau{He}{4}{}$ and $\noyau{Li}{7}{}$ fragments for the carbon and titanium targets. The two highlighted regions correspond to the experimental energy thresholds (gray shading, cf.~Table \ref{tab:EThr}) and to the energy range of the beam when passing through the target (red shading).}
\label{fig:EDist}
\end{figure}

Regarding the double differential fragmentation cross sections, Fig.~\ref{fig:EDist} represents the energy distributions of $\noyau{He}{4}{}$ and $\noyau{Li}{7}{}$ obtained for the carbon and titanium targets. The highlighted region correspond to the energy range of the $\noyau{C}{12}{}$ projectile when passing through the target. The fragmentation processes occur therefore in this energy range. The distributions are peaked at the beam energy. This observation is consistent with the results of the previous experiment in which it was shown that the energy distributions were peaked at an energy close to the beam energy for the smallest angles.

\subsection{Comparisons with data at larger angles}

\begin{figure}[!ht]
\includegraphics[width=0.5\linewidth]{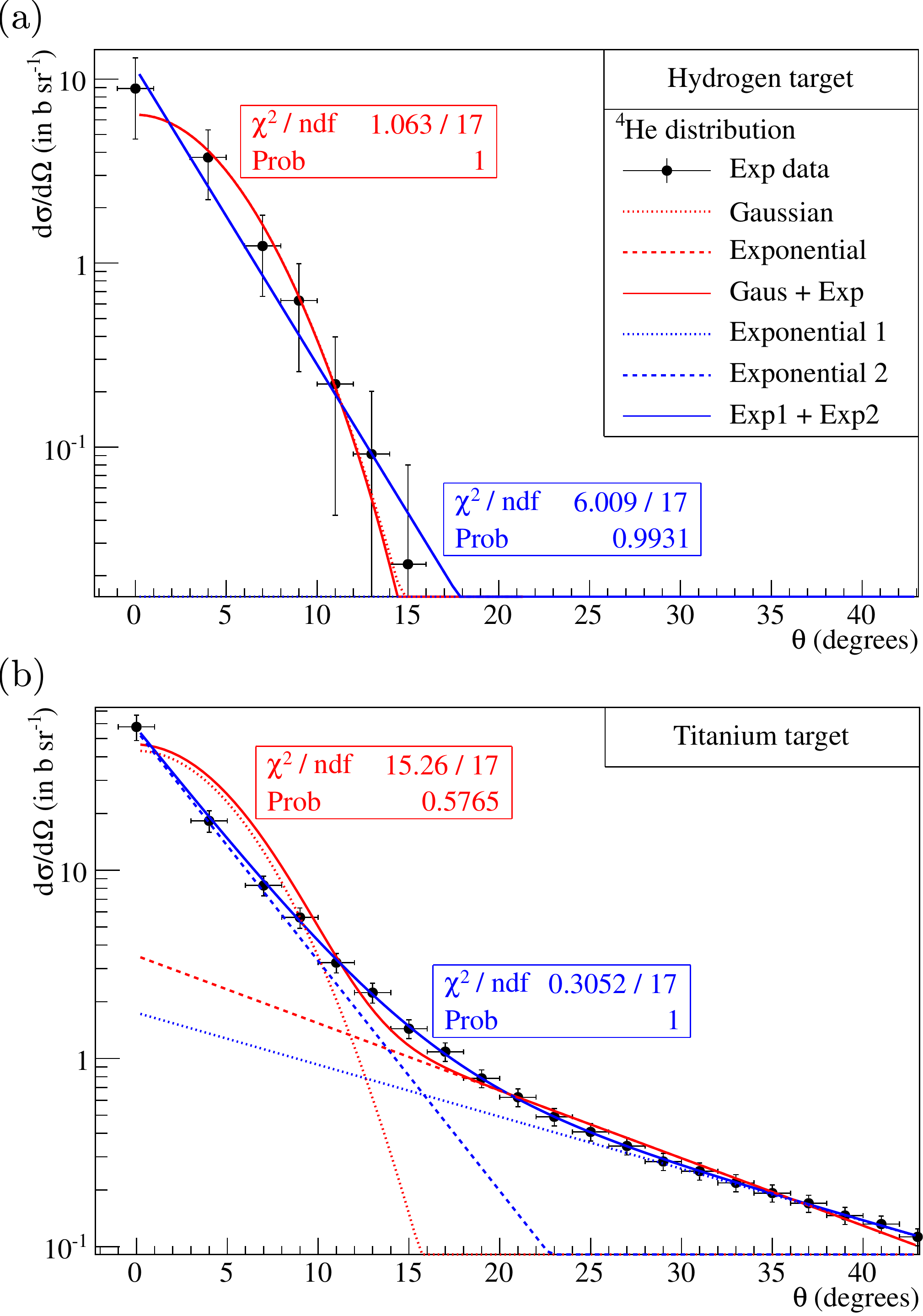}
\caption{(Color online) $\noyau{He}{4}{}$ angular distributions for the hydrogen (a) and titanium (b) targets. The distributions have been fitted with two functions. The first one (red lines) resulting from the sum of a gaussian and an exponential functions and the second one (blue lines) resulting from the sum of two exponential functions.}
\label{fig:GEvsEE}
\end{figure}

The fragmentation differential cross section at zero degree have been presented. In this last part, comparisons will be done between these zero degree data and the one obtained at larger angles. In our previous work, we show that the angular distributions were well fitted by a function resulting of the sum of a gaussian (for forward angles) and an exponential (for large angles) function (in red on Fig.~\ref{fig:GEvsEE}). This ability to fit the angular distributions with an analytical function allows to obtain the production cross section of each isotope by integrating this fitted function over the whole solid angle. However, the smallest angle available was 4$^\circ$ and the integral of the distribution is very dependent on the zero degree value.

By taking into account the zero degree value in the fit of the distribution, the gaussian and exponential description of the distributions is not valid anymore (cf.~Fig.~\ref{fig:GEvsEE}). The gaussian description at small angles seems still valid for light targets like the hydrogen target (Fig.~\ref{fig:GEvsEE}(a)), but not for heavier targets like the titanium one (Fig.~\ref{fig:GEvsEE}(b)). It appears that for heavy targets, the behavior at small angles is better reproduced by replacing the gaussian function by an exponential function (in blue on Fig.~\ref{fig:GEvsEE}).

\begin{figure*}[!ht]
\includegraphics[width=1\linewidth]{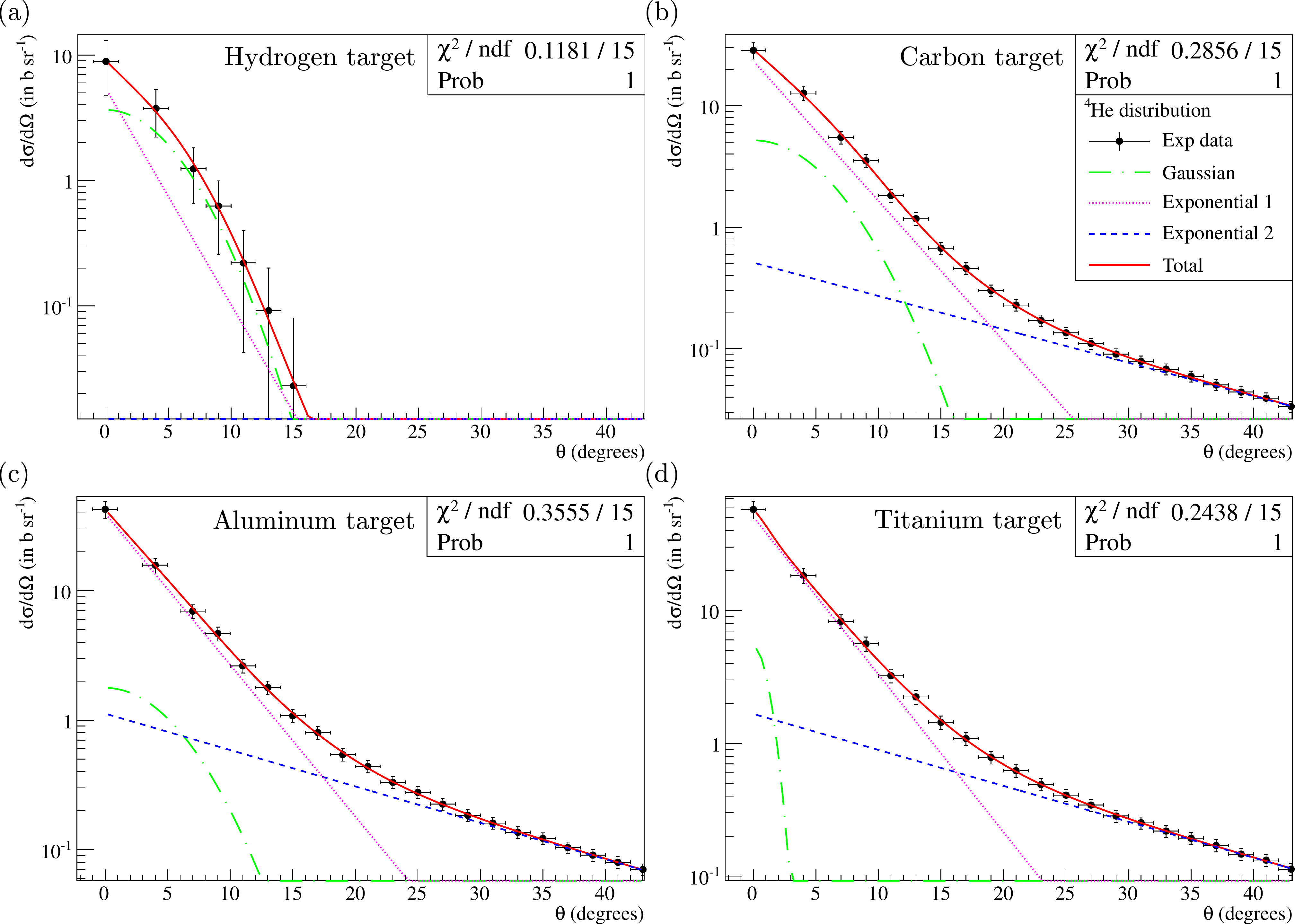}
\caption{(Color online) $\noyau{He}{4}{}$ angular distributions for the hydrogen (a), carbon (b), aluminum (c) and titanium (d) targets. The distributions have been fitted with a function resulting of the sum of a gaussian and two exponential functions.}
\label{fig:GEE_AllTarg}
\end{figure*}

Due to this gaussian/exponential behavior at small angles depending on the target mass, new fits have been performed with a function resulting from the sum of a gaussian and two exponential functions. These new fitting functions are represented in Fig.~\ref{fig:GEE_AllTarg}, on the $\noyau{He}{4}{}$ distributions for the hydrogen (a), carbon (b), aluminum (c) and titanium (d) targets. The fit results are very close to data with $\chi^2$ values of about 0.2. These values lower than one are due to the experimental uncertainties which are probably overestimated. The global errors are dominated by systematic errors (beam monitor and solid angle of the telescopes) which have been voluntarily determined in a conservative way. 

It can also be observed on Fig.~\ref{fig:GEE_AllTarg} that the heavier the target, the smaller is the gaussian contribution and consequently, the more important are the two exponential contributions. The same behavior is observed for the other isotopes.

\begin{figure*}[!ht]
\includegraphics[width=1\linewidth]{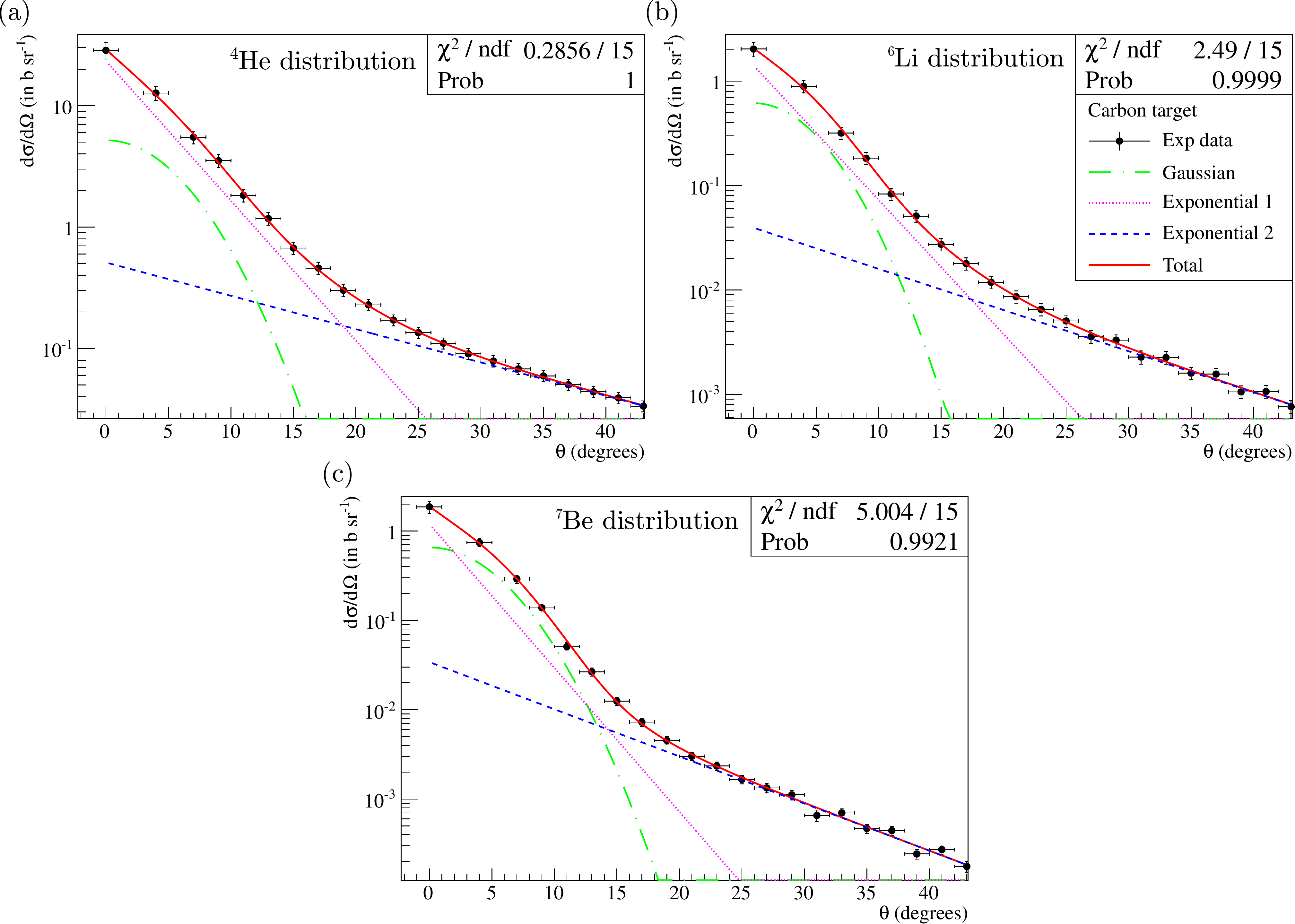}
\caption{(Color online) Angular distributions of $\noyau{He}{4}{}$ (a), $\noyau{Li}{6}{}$ (b) and $\noyau{Be}{7}{}$ (c) obtained for the carbon target. The distributions have been fitted with a distribution resulting of the sum of a gaussian and two exponential functions.}
\label{fig:GEE_AllFrag}
\end{figure*}

The same fitting functions are represented on Fig.~\ref{fig:GEE_AllFrag}, for the distributions of $\noyau{He}{4}{}$ (a), $\noyau{Li}{6}{}$ (b) and $\noyau{Be}{7}{}$ (c), obtained for the carbon target. The fit results are still very close to data and we can see that the gaussian contribution is more important for heavier produced fragments. This observation has also been verified for the other targets.

Even if we are not able to give a physical explanation to this behavior, this allows us to obtain a very good data reproduction and consequently, a better estimation of the production cross sections. Nevertheless, it appears that the production cross section obtained by integrating this new function are not so different from the one obtain with the previous fitting functions. Table~\ref{tab:ProdXSec} represents the new production cross sections, compared to the one obtained by fitting the distributions with a gaussian and exponential distribution, without the zero degree measurements, as published in Dudouet {\it et al.}~\cite{Dudouet13b} (with data rectified from the beam monitor calibration correction).

In most cases, even without the zero degree values and using a gaussian and exponential fit, the relative error on the production cross sections is on average of 4\%.

\begin{table*}[!ht]
\begin{center}
\begin{tabular}{l c c c}
\hline
\hline
\multicolumn{1}{l}{\multirow{2}{3cm}{Target}} & \multicolumn{3}{c}{$\sigma$ (b)}\\
\cline{2-4}
	& 	\multirow{1}{3cm}{\centering $\noyau{He}{4}{}$}	& 	\multirow{1}{3cm}{\centering $\noyau{Li}{7}{}$}	& 	\multirow{1}{3cm}{\centering $\noyau{Be}{7}{}$}	\\
\hline
H (Gaus+Exp+Exp) 	& 2.01(0.27)\ee{-1}	& 7.9(5.2)\ee{-3}	& 1.47(0.75)\ee{-2}		\\

H (Gaus+Exp)		& 1.80(0.61)\ee{-1}	& 7.3(4.2)\ee{-3}	& 1.44(0.34)\ee{-2}		\\
\hline
Rel err (in \%)		& 10.4			&  7.6			& 2.0				\\
\hline
\hline

C (Gaus+Exp+Exp) 	& 1.09(0.38)		& 5.45(0.84)\ee{-2} 	& 4.55(0.72)\ee{-2} 		\\

C (Gaus+Exp)		& 1.08(0.11)		& 5.39(0.78)\ee{-2}	& 4.49(0.65)\ee{-2}		\\
\hline
Rel err (in \%)		& 1.0			& 1.1			& 1.3				\\
\hline
\hline

Al (Gaus+Exp+Exp) 	& 1.58(0.14)		& 7.9(1.1)\ee{-2} 	&  5.99(0.87)\ee{-2}		\\

Al (Gaus+Exp)		& 1.53(0.22)		& 7.77(0.95)\ee{-2}	& 5.88(0.76)\ee{-2}		\\
\hline
Rel err (in \%)		& 3.2			& 1.3			& 1.8				\\
\hline
\hline

Ti (Gaus+Exp+Exp) 	& 2.06(0.31)		& 1.01(0.13)\ee{-1} 	& 7.0(1.1)\ee{-2} 		\\

Ti (Exp+Gaus)		& 1.96(0.30)		& 9.6(1.1)\ee{-2}	& 6.68(0.83)\ee{-2}		\\
\hline
Rel err (in \%)		& 4.9			& 5.0			& 4.3				\\
\hline
\hline
\end{tabular}
\end{center}
\caption{New production cross sections obtained by fitting the distributions with a gaussian and two exponential functions. They are compared to the previous ones obtained by fitting the distributions with a gaussian and an exponential function, without including the zero degree measurements, as published in Dudouet {\it et al.}~\cite{Dudouet13b} (but with data rectified from the beam monitor calibration correction).}
\label{tab:ProdXSec}
\end{table*}

\section{Conclusions}

A first experiment was performed on May 2011 by our collaboration and allowed to obtain double differential fragmentation cross sections of 95~MeV/A $\noyau{C}{12}{}$ on different thin targets at angles from 4 to 43$^\circ$. The results obtained from the analysis of the new experiment presented in this work permit to complete these data with zero degree measurements on hydrogen, carbon, aluminum and titanium targets. 

A precise analysis of the beam monitor revealed a non linearity of the plastic scintillator when five or more incident ions per beam bunch are detected. By taking into account this information, the beam monitor calibration of the first experiment has been corrected. As a result, the data previously published in Dudouet {\it et al.}~\cite{Dudouet13b} have been corrected of about 5\%. These new corrected data, as the new zero degree values, have been updated and are available in free access on the web-site: \url{http://hadrontherapy-data.in2p3.fr}.

As a first result of this experiment, a telescope that was located at an angle of 9$^\circ$ with respect to the beam direction permitted to validate the results obtained in the first experiment within a 3\% agreement.

Regarding the zero degree measurements, a specific analysis was required to be able to distinguish the events due to the beam interactions in the detectors regarding the fragmentation events that have taken place in the target. Thanks to this analysis, the differential cross section have been obtained, for fragments of Z=2 to Z=5 and for the most produced isotopes of these Z values. GEANT4 simulations were performed using the nuclear model INCL++ to estimate the systematic errors mainly due to mismatch identifications from nuclear interactions in the detectors. These zero degree cross sections have been measured within a 10 to 15\% accuracy, including statistical and systematic errors.

Moreover, the angular distributions are strongly dominated by the statistics at small angles, especially at zero degree. Regarding the energy distributions of these zero degree emitted fragments, the latter are mostly emitted at the beam velocity, due to the projectile fragmentation.

Finally, a previous analysis shows that the angular distributions were well represented by a function resulting of the sum of a gaussian and an exponential function~\cite{Dudouet13b}. Nevertheless, by applying the same fit on the data including the zero degree values, we have shown that this gaussian and exponential representation is not valid anymore at zero degree. In order to obtain a more precise estimation of the production cross sections, the distributions have been fitted with a new function, resulting from the sum of a gaussian and two exponential functions, one exponential function being dominant at small angles and the other being dominant at large angles. The total fragment production cross sections obtained from the previous fit and the new one differ on average by 4\%. 

To fulfill the study on the all therapeutic energies, experiments at 50~MeV/A are expected at GANIL in the following years and systematic measurements up to 400 MeV/A are planned at ARCHADE (resource center for hadrontherapy in Caen) from 2020. 

\clearpage

\bibliographystyle{unsrt}
\bibliography{ZeroDeg}

\end{document}